# Resistivity of Mn$_{1-x}$Fe$_x$Si single crystals: Evidence for quantum critical behavior

**Christoph Meingast,[1] Qin Zhang,[1*] Thomas Wolf[1], Frédéric Hardy,[1] Kai Grube,[1] William Knafo,[1,2+] Peter Adelmann,[1] Peter Schweiss[1], and Hilbert v. Löhneysen[1,2]**

[1] Forschungszentrum Karlsruhe, Institut für Festkörperphysik, PO Box 3640, 76021 Karlsruhe.

[2] Physikalisches Institut, Universität Karlsruhe, 76128 Karlsruhe.

**Abstract**    Resistivity measurements have been made on Mn$_{1-x}$Fe$_x$Si single crystals between 2 K and 300 K for $x$ = 0, 0.05, 0.08, 0.12 and 0.15.  Fe doping is found to depress the magnetic ordering temperature from 30 K for $x$ = 0 to below 2 K for $x$ = 0.15.  Although Fe doping results in a large increase of the low-temperature residual resistivity, the temperature dependence of the resistivity above the magnetic transition remains practically unaffected by increasing Fe content.  An analysis of the temperature derivative of the resistivity provides strong evidence for the existence of a non-Fermi-liquid ground state near x = 0.15 and thus for a quantum critical point tuned by Fe content.

## Introduction

MnSi is thought to be a good example of a weak itinerant ferromagnet [1, 2]. Actually, due to the lack of space inversion symmetry of the B20 crystal structure and the resulting Dzyaloshinski-Moriya spin-orbit interaction, MnSi[1] possesses a long-wavelength (180 Angstroms) helical rotation of the magnetization in zero magnetic field [3].  The magnetic ground state of MnSi is thus better described as incommensurate antiferromagnetic. However, above a field of roughly 0.6 T, the helical order is quenched and the systems then behaves like a typical weak itinerant ferromagnet [3]. The smallest energy scale in MnSi is the crystal-field interaction, which pins the helices along one of the 111 crystallographic directions. The interest in MnSi has recently been renewed by the discovery of a large region of non-Fermi liquid behavior in the resistivity with a $T^{3/2}$ power law above the critical pressure (~ 15 kbar) where the long-range magnetic order disappears [4]. Usually non-Fermi-liquid behavior is limited to a small region around a quantum critical point [5,6], and the reason for this large region of non-Fermi liquid behavior remains unknown [7].

Here we study the resistivity of $Mn_{1-x}Fe_xSi$. Previous work has shown that the magnetic order is suppressed with increasing Fe content [8,9], and it is thus of strong interest to see if a concentration-tuned quantum critical point occurs in $Mn_{1-x}Fe_xSi$. Resistivity measurements provide a simple, but quite powerful, means to investigate such transitions. In particular, there appears to be a very strong correlation of the temperature derivative of the resistivity with the magnetic/electronic heat capacity in strongly correlated systems [10,11]. A more detailed report of these data, as well as magnetization, specific heat and thermal expansion will be published elsewhere [12].

## Experimental

Single crystals of $Mn_{1-x}Fe_xSi$ have been grown by the vertical Bridgman method in conical $Al_2O_3$ crucibles. Stoichiometric mixtures of Mn (Cerac, 4N), Si (Alfa, 6N), and Fe (Fluka, 2N5) were heated to 1460 °C at a rate of 250 °C/h, and after a holding time of 1 h, cooled down to 1150 °C at rates between 1- 2 °C/h. At the beginning the growth chamber was evacuated overnight to pressures < $10^{-5}$ mbar and then, at temperatures > 1000 °C, backfilled to 1100 mbar with 6N argon. To further reduce the oxygen content in the growth atmosphere the $Al_2O_3$ crucible was packed into Zr powders as a getter material. Crystals with typical dimensions of 10 x 1 x 0.5 mm$^3$ were cut with a wire saw and mechanically polished to the final dimensions. Some of the crystals had small pores, which made an accurate determination of the absolute resistivity problematic (see below). The resistivity was measured using a four-points ac measurement technique in a PPMS (Physical Property Measurement System) from Quantum Design. The contacts were made using silver paint and gold (or platinum) wires.

## Results and Discussion

Fig. 1 shows the raw resistivity data of $Mn_{1-x}Fe_xSi$ between 2 K and 300 K for $x = 0, 0.05, 0.08, 0.12$ and 0.15. The resistivity of pure MnSi is very similar to previous measurements [13-15]. The present residual resistivity ratio (RRR), $\rho(300$ K$)/\rho(2$ K$) = 21.5$, is smaller than the highest RRR (up to 100 – 200) reported in the literature [13-15]. Interestingly, we also observed higher RRR ratios for some crystals, which however had broader magnetic transitions. We do not have an explanation for this behavior, and here we present only the data of the crystal with a sharp magnetic phase transitions and a lower RRR. The residual resistivity increases strongly with x. However, the general shape of the resistivity curves above $T_N$ is hardly affected by an increase of the Fe content. The curves for $x = 0.05$ and 0.08 have somewhat steeper slopes, however this may be due to an uncertainty in the absolute values due to the previously mentioned pores in some of the samples. In order to have a better comparison of the shape of the resistivity curves, we have normalized the resistivity curves at 300 K (see Fig. 2). Now it is clearly seen that, besides the increase of the residual resistivity, the curves all have very similar shapes above $T_N$.

In oder to study the magnetic phase transition at $T_N$ in more detail, we calculated the temperature derivative of the resistivity, $d\rho(T)/dT$, is plotted versus $T$ in Fig. 3. The data have been normalized at 100 K. The phase transition is now clearly seen and strongly resembles the behavior of the specific heat and thermal expansion [12,15]. We note that the transition takes place in two stages; coming from low temperatures, $d\rho(T)/dT$ first has a very sharp peak, followed by a broadened hump just above the peak. We believe that the sharp peak signals the real phase transition, i.e. the point where helical order is established with long-range phase coherence throughout the crystal. This transition is weakly first-order. The broad hump above this peak, on the other hand, is probably due to the establishment of ferromagnetic-like correlations. This is because, with the application of an applied magnetic field, the sharp peak decreases in temperature, whereas the broad hump becomes broader and moves to higher temperatures, as would be expected for a simple ferromagnetic transition [12, 15]. With increasing Fe content, the transitions remain remarkably sharp for $x = 0.05$ and 0.08. For $x = 0.12$, the transition is somewhat broadened, and the sharp first-order transition is no longer observed. For $x = 0.15$, we see no clear sign of a transition; $d\rho(T)/dT$ first increases with increasing temperature and then goes over a broad maximum at roughly 4 K.

In order to further analyze these resistivity data, it is very useful to look at $d\rho(T)/dT$ divided by temperature (see Fig. 4). $(1/T) \cdot d\rho(T)/dT$ is analogous to the magnetic/electronic specific heat divided by $T$, i.e. $C_p/T$ [10,11,15]. In Fig. 4 we see that $(1/T) \cdot d\rho(T)/dT$ becomes constant at low T for pure MnSi, as would be expected for a Fermi-liquid like state with a constant $C_p/T$. For increasing Fe content x, $(1/T) \cdot d\rho(T)/dT$ is no longer constant at low temperature, but rather approaches a nearly logarithmic temperature dependence, indicating the approach to a non Fermi-liquid state. For x = 0.15, $(1/T) \cdot d\rho(T)/dT$ increases smoothly to the lowest measured temperature (2 K) and shows no sign of a finite-temperature phase transition. The present data thus provide strong evidence that Mn$_{1-x}$Fe$_x$Si approaches a magnetic quantum critical point for $x$ close to 0.15. This nearly logarithmic divergence is in contrast to the $T^{-0.5}$ divergence expected from the $\rho(T) \sim T^{3/2}$ power law found for pure MnSi in the region above the critical pressure [4]. Here, it is worth recalling that the transition under pressure to a non-magnetic state in pure MnSi is of first order above a pressure $p^*$ close to the critical pressure, whereas the present data point more to a second-order transition with Fe substitution. The difference in critical behavior may, thus, be due to the different tuning parameters (pressure versus composition) employed to approach the quantum critical point.

## Conclusions

In summary, resistivity measurements of $Mn_{1-x}Fe_xSi$ single crystals between 2 K and 300 K nicely demonstrate that Fe doping depresses the magnetic ordering temperature from 30 K for $x = 0$ to below 2 K for $x = 0.15$. A detailed analysis of the temperature derivative of the resistivity provides strong evidence for the existence of a non-Fermi-liquid ground state near $x = 0.15$. This makes $Mn_{1-x}Fe_xSi$ a very interesting system for further studies. Of special interest is the comparison with the behavior of pressure-tuned pure MnSi. Preliminary thermal expansion and specific heat data of $Mn_{1-x}Fe_xSi$ are in accord with the present resistivity data and also provide evidence for a quantum critical point near x = 0.15 [12]. Together these data show that the Grüneisen parameter, which is just the ratio of the thermal expansion coefficient and the specific heat, exhibits a divergence and a sign change, as is typically expected close to a pressure-tuned quantum phase transition [16].

**Acknowledgments** We would like to acknowledge useful discussion with Daniel Lamago and Dmitry Reznik and to thank Severin Adandogou and Doris Ernst for technical assistance.

References

1. T. Moriya, Spin Fluctuations in Itinerant Electron Magnetism, vol. 56 of Solid-State Sciences (Springer, Berlin, 1985).
2. G. G. Lonzarich and L. Taillefer, J. Phys. **C 18**, 4339 (1985).
3. Y. Ishikawa et al., Phys. Rev. **B 16**, 4956 (1977).
4. N. Doiron-Leyraud et al., Nature **425**, 595 (2003).
5. G. R. Stewart, Rev. Mod. Phys. **73**, 797 (2001).
6. H. v. Löhneysen et al., Rev. Mod. Phys. **79**, 1015 (2007).
7. C. Pfleiderer et al., Science **316**, 1871 (2007).

8. N. Manyala et al., Nature **404**, 581 - 584 (2000).

9. D. Shinoda, phys. stat. sol. (a) **11**, 129 (1972).

10. M. E. Fisher and J.S. Langer, Phys. Rev. Lett. **20**, 665 (1968).

11. N. Sakamoto et al., Phys. Rev **B 69**, 092401 (2004).

12. Q. Zhang et al., (unpublished).

13. C. Pfleiderer et al., Phys. Rev. **B55**, 8330 (1997).

14. F. P. Mena et al., Phys. Rev. **B 67**, 241101 (2003).

15. S. M. Stishov et al., Phys. Rev. **B 76**, 052405 (2007).

16.  M. Garst and A. Rosch, Phys. Rev. **B 72**, 205129 (2005).

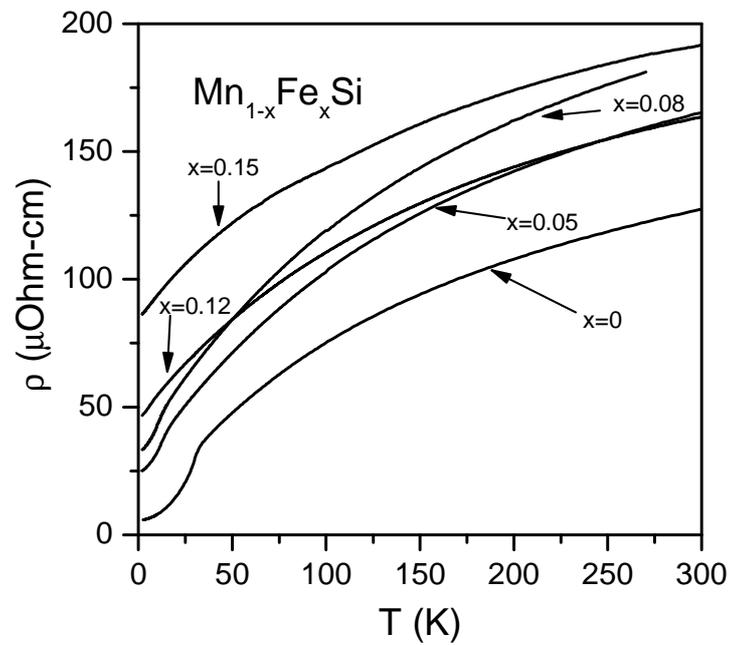

**Fig. 1.** Resistivity versus temperature of $Mn_{1-x}Fe_xSi$.

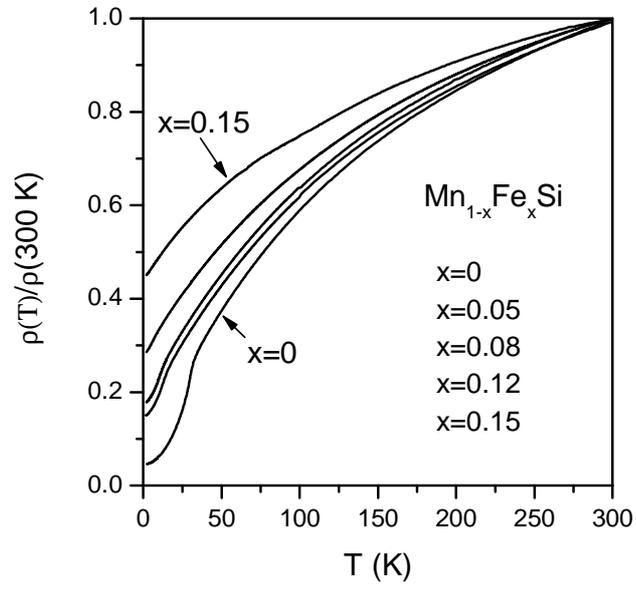

**Fig. 2.** Normalized resistivity, $\rho$(T)/$\rho$(300 K), versus temperature of Mn$_{1-x}$Fe$_x$Si..

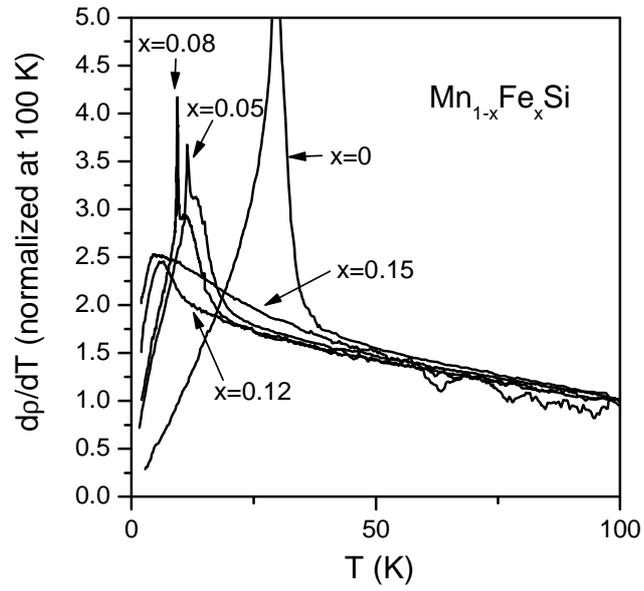

**Fig. 3.** Normalized (at 100 K) temperature derivative of the resistivity $\rho$(T)/$d$T of Mn$_{1-x}$Fe$_x$Si. The magnetic phase transition is clearly seen for all Fe concentrations except for x = 0.15..

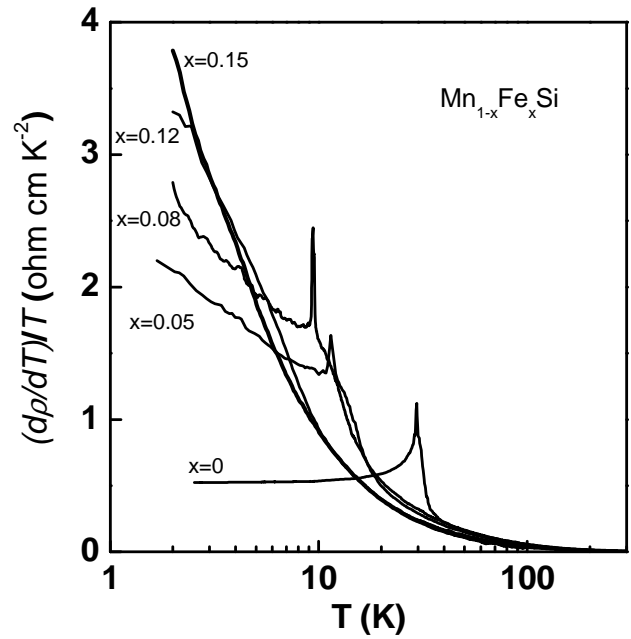

**Fig. 4.** $1/T \cdot d\rho(T)/dT$ versus temperature on a log(T) scale for Mn$_{1-x}$Fe$_x$Si. $1/T \cdot d\rho(T)/dT$ is constant at low temperature for pure MnSi indicative of a Fermi-liquid state. For x = 0.16, on the other hand, $1/T \cdot d\rho(T)/dT$ diverges nearly logarithmically implying a non-Fermi-liquid ground state.